\documentclass[preprint,showpacs,preprintnumbers,amsmath,amssymb,showkeys]{revtex4}
\usepackage{times, amsmath, latexsym, amssymb, graphicx,amsthm}

\newcommand{\ddx}{\frac{\partial}{\partial x}}

\newcommand{\ddy}{\frac{\partial}{\partial y}}

\newcommand{\dtdy}{\frac{\partial^2}{\partial y^2}}

\newcommand{\rme}{\mathrm{e}}

\begin{document}
\title{Leray-$\alpha$ model and transition to turbulence in rough-wall boundary layers}
\author{Alexey Cheskidov}
\email{acheskid@umich.edu}
\affiliation{Department of Mathematics,
University of Michigan, Ann Arbor, Michigan 48109}
\author{Diana Ma}
\email{dkma@umich.edu}
\affiliation{Department of Mechanical Engineering
University of Michigan, Ann Arbor, Michigan 48109}
\date{\today}

\begin{abstract}
In the present study, we use a fifth-order ordinary differential equation, a generalization
of the Blasius equation derived from the Leray-$\alpha$ model,
to examine the transition to turbulence in rough-wall boundary-layer flows. 
This equation, combined with a weaker formulation of the
von K\'{a}rm\'{a}n log law modified to include the effects of
surface roughness, provides a family of turbulent velocity profiles
with two free parameters: the momentum thickness based Reynolds
number and the roughness function. As a result, a 
skin-friction correlation is obtained and predictions of the transitional Reynolds
numbers and maximal skin-friction values are made based on the roughness
function.
\end{abstract}



\maketitle

\section{Introduction}
Wall-bounded turbulent flows continue to be significant in both the
natural environment and in engineering applications. As such, the
turbulence community has persistently endeavored to describe the
fluid dynamics in such flows \cite{S}. In particular, the
determination of skin-friction (or wall shear stress) has been of
interest to many researchers.

For smooth-wall turbulent boundary layers, both direct and indirect
techniques have been developed to determine skin-friction. These
include the use of the momentum integral equation, correlations
based on pressure measurements at the surface, and fitting mean
velocity profiles based on a defect or power law \cite{Bergstrom}.

However, in the rough-wall case \cite{G,J}, direct measurement of skin-friction
is often difficult. Indirect methods have included using friction
velocity, $u_{\tau}$, to estimate skin-friction. The modified
Clauser method, for instance,
has been used to approximate $u_{\tau}$ by fitting a logarithmic velocity profile
to experimental data. However, as discussed by Acharya and Escudier \cite{Acharya}, this
technique is subject to large uncertainties because the degrees of
freedom are increased from one ($c_{f}$) for a smooth surface to
three ($c_{f}$, $\epsilon$, $\Delta u/u_{\tau}$), where $c_{f}$ is the skin-friction
coefficient, $\epsilon$ is  the error in origin,
and $\Delta u/u_{\tau}$ is the roughness function.
Alternative indirect
techniques include determining $u_{\tau}$ using a velocity defect
law, or power law formulations. Other correlations such as Bergstrom
et al.'s skin-friction correlation with the ratio of the
displacement and boundary-layer thicknesses have also been suggested
\cite{Bergstrom}.

In \cite{Cheskidov2005}, a new theoretical method to derive a skin-friction
correlation for smooth-wall turbulent boundary layers was presented.
In this note we adopt it to rough-wall boundary-layer flows using
the boundary-layer approximation of the Leray-$\alpha$ model of
turbulence \cite{CHOT} and the von K\'{a}rm\'{a}n log law  for rough walls
\cite{Clauser1954, Clauser1956,N}.
The benefit of this approach is that it leads to a prediction of the critical Reynolds number based
on momentum thickness $R^{\mathrm{crit}}_\theta$, the minimal Reynolds
number where the transition to turbulence may occur.
More precisely, we obtain the following dependence:
\[
R_\theta^\mathrm{crit} = -51.8 \Delta u/u_{\tau} + 365.5.
\]
Since in a turbulent boundary layer, the skin-friction coefficients attains its maximum $c_{f}^\mathrm{max}$ at $R_\theta = R_\theta^\mathrm{crit}$, we also obtain the
the following relation:
\[
c_{f}^\mathrm{max} = 0.0063e^{0.1861 \Delta u/u_{\tau}}.
\]

\section{Boundary-layer approximation of the Leray-$\alpha$ model}
Proposed as a closure scheme for the Reynolds equations \cite{CHOT}, the Leray-$\alpha$ model is written as
\begin{equation} \label{leray-introduction}
\left\{
\begin{aligned}
&\frac{\partial}{\partial t} \mathbf{v} + (\mathbf{u} \cdot
\nabla)\mathbf{v} =
\nu\Delta \mathbf{v} - \nabla p + f\\
&\nabla \cdot \mathbf{u} = 0\\
&\mathbf{v}=\mathbf{u}- \alpha^2 \Delta \mathbf{u},
\end{aligned}
\right.
\end{equation}
where $\mathbf{u}$ is the averaged physical velocity of the flow,
$p$ is the averaged pressure, $f$ is a force, and $\nu>0$ is the
viscosity.
The filter length scale $\alpha$ represents the averaged
size of the Lagrangian fluctuations and is
considered as a parameter of the flow. More specifically, we assume
that $\alpha$ changes along the streamlines in the boundary layer, and is
proportional to the thickness of the boundary layer (see \cite{C}).
Inspired by the Navier-Stokes-$\alpha$ model (see \cite{CFHOTW3} and
references therein), this model compared successfully with experimental
data from turbulent channel, pipe, and boundary-layer flows
for a wide range of Reynolds numbers.

We recall the boundary-layer approximation of the Leray-$\alpha$ model
(see \cite{Cheskidov2005}).
In the case of a zero-pressure gradient, consider a two-dimensional
flow across a flat surface. Let
$x$ be the coordinate along the surface, $y$ the coordinate normal
to the surface, and $\mathbf{u}=(u,v)$ the velocity of the flow.
Assuming that $\alpha$ is a function of $x$ only, normalizing
variables, and neglecting terms that are small near the boundary
(see \cite{C2}), we arrive at a Prandtl-like
boundary-layer approximation of the 2D Leray-$\alpha$ model:
\begin{equation} \label{eq:p-l}
\left\{
\begin{aligned}
&u\ddx w + v\ddy w= \dtdy w  \\
&v (x,y)= -\int_0^y \ddx u(x,z) \, dz\\
&w=u-\alpha^2 \dtdy u,
\end{aligned}
\right.
\end{equation}
where $(u,v)$ are the components of the averaged velocity, $p$ the
averaged pressure, and $w=\left(1-\alpha^2\dtdy\right)u$.

The physical (non-slip) boundary conditions are $u(x,0)=v(x,0)=0$,
and $(u(x,y),v(x,y)) \to (1,0)$ as $y \to \infty$. Looking for self-similar solutions
to this system of the form
\[
u(x,y) =\frac{1}{\beta^2} h'(\xi/\beta), \qquad \alpha(x)=\beta \delta(x), \qquad \xi
=\frac{y}{\delta(x)},
\]
with $\delta(x)=\sqrt{x}$, we reduce \eqref{eq:p-l} to
the following generalization of the celebrated Blasius equation:
\begin{equation} \label{eq:systemintro} 
m''' + \frac{1}{2}hm''=0, \qquad m=h-h''.
\end{equation}
The physical boundary conditions for \eqref{eq:systemintro} are
$h(0)=h'(0)=0$  and $h'(\xi) \rightarrow \beta^2$ as $\xi \rightarrow
\infty$.
This equation describes horizontal velocity profiles $\{h'(\cdot)\}$
in transitional and turbulent boundary layers with zero pressure
gradients. In \cite{C2} it was proved that the above boundary value problem
has a two parameter family of solutions with the
parameters being
\[
a:= h''(0), \qquad b:=h'''(0).
\]

From the derivation of \eqref{eq:systemintro} it follows that the averaged 
horizontal velocity profiles $u$ for a fixed horizontal coordinate $x_0$ is
modeled by
\begin{equation} \label{eq:uyconn}
u(x_0,y)=\frac{u_{\rme}}{\beta^2} h'\left(\frac{y}{\beta
\sqrt{l_{\rme}l}}\right),
\end{equation}
where $y$ is the vertical coordinate, $u_{\rme}$ is the
horizontal velocity of the external flow, $h$ is a solution to
(\ref{eq:systemintro}), $\beta = (\lim_{y \to \infty} h'(y))^{1/2}$,
$l$ is a local length scale that has to be determined, $R_l = u_{\rme}l/\nu$, and
$l_{\rme}$ is the external length scale $l_{\rme}=\nu / u_{\rme}$.

We now normalize quantities into wall coordinates
\[ y^+=\frac{u_\tau y}{\nu}, \qquad
u^+=\frac{u}{u_\tau}
\]
where
\[
u_\tau=\sqrt{\frac{1}{\rho}\tau}= \sqrt{\nu\frac{\partial
u}{\partial y}\Big|_{y=0}} \, ,
\]
and $\tau$ is the shear stress at the wall.
Writing (\ref{eq:uyconn}) in wall units, a three-parameter family of
velocity profiles is obtained $u^+_{a,b,l}(\cdot)$:
\begin{equation} \label{eq:threeparam}
u^+_{a,b,l}(y^+)=\frac{R_l^{1/4}}{\sqrt{a \beta}} h' \left(
\frac{y^+ \sqrt{\beta}}{R_l^{1/4} \sqrt{a}}  \right).
\end{equation}

\section{Skin-friction correlation}
A critical step in deriving a skin-friction correlation is reducing a
three-parameter family of the velocity profiles \eqref{eq:threeparam}
to a two-parameter family of turbulent velocity profiles, which is achieved
with a help of the von K\'{a}rm\'{a}n log law.

For smooth surfaces, the mean velocity profile for the inner region
is commonly approximated with the von K\'{a}rm\'{a}n log law:
\[
u^+ = \frac{1}{\kappa} \ln y^+  + B,
\]
where the von K\'{a}rm\'{a}n constant, $\kappa \approx 0.4$, and $B
\approx 5$, are empirically determined constants.
In the rough-wall case, the effects of uniform roughness are
confined to the inner region, and are accounted for by modifying the
semi-logarithmic part of the mean velocity profile. More
specifically, Clauser \cite{Clauser1954, Clauser1956} showed that
the semi-logarithmic region is displaced downward by an amount
$\Delta u/u_{\tau}$. This amount of downward shift is commonly
referred to as the roughness function, and represents the velocity
defect from the standard velocity distribution over a smooth wall,
and indicates the additional wall shear stress due to the roughness.
Accounting for the roughness effect, the log law can then be written
as
\begin{equation} \label{e:loglaw}
u^+ = \frac{1}{\kappa} \ln \left[\frac{(y +
\varepsilon)u_{\tau}}{\nu}\right] + B - \Delta B,
\end{equation}
where $\varepsilon$ is the shift at the origin for the rough wall,
$y$ is measured from the top of the roughness element, and $\Delta B
= \Delta u/u_{\tau}$. The values of $\varepsilon$ and
$\Delta B$ are determined by matching experimental velocity profiles
with \eqref{e:loglaw}.

\begin{figure}
\center
\includegraphics[width=\textwidth]{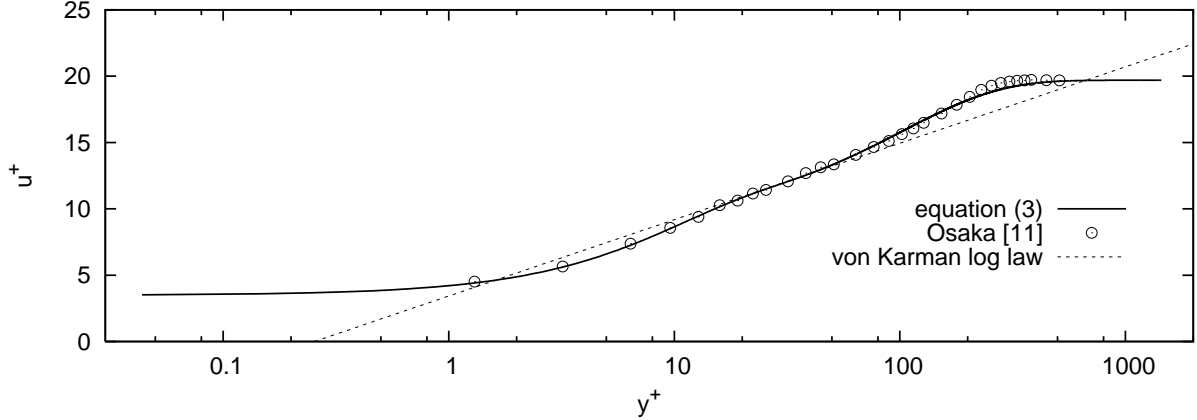}
\caption{Velocity profile for $R_\theta=700$, $\Delta B = 1.57$.} 
\label{f1}
\end{figure}

We obtain turbulent velocity profiles by subjecting profiles \eqref{eq:threeparam}
to three conditions of a weaker formulation of the von K\'{a}rm\'{a}n
log law:
\begin{enumerate}
\item \label{first} A turbulent velocity profile $u_\mathrm{t}^+(y^+)$ has
$3$ inflection points in logarithmic coordinates.
\item \label{second} The middle inflection point of $u_\mathrm{t}^+(y^+)$
lies on the line
\begin{equation} \label{eq:logline}
u^+=\frac{1}{\kappa}\ln y^+  + B - \Delta B.
\end{equation}
\item \label{third} The line~(\ref{eq:logline}) is tangent to
$u_\mathrm{t}^+(y^+)$ at the middle inflection point.
\end{enumerate}
Let us fix a Reynolds number based on momentum thickness
\begin{equation} \label{condition2}
R_\theta = \int_0^{\infty} u^+\left( 1-
\frac{u^+}{u^+(\infty)}\right) \,dy^+.
\end{equation}
Then \eqref{condition2} and conditions (ii) and (iii) determine
all three parameters $a$, $b$, and $l$ in $\eqref{eq:threeparam}$.
Therefore, conditions (ii) and (iii) reduce $\eqref{eq:threeparam}$
down to a two-parameter family of turbulent profiles $\{u^+_{R_\theta,\Delta B}\}$.
Fig.~\ref{f1} shows the velocity profile for $R_\theta = 700$,
$\Delta B = 1.57$ and corresponding experimental data of
Osaka et al. \cite{Osaka} for d-type roughness.

Note that the skin-friction coefficient is uniquely determined by a velocity
profile $c_f = 2/u^+_{R_\theta,\Delta B}(\infty)^2$. Therefore, the skin-friction coefficient is now a function of $R_\theta$ and $\Delta B$:
\begin{equation} \label{eq:cflaw}
c_f = f(R_\theta, \Delta B).
\end{equation}

\section{The roughness function and transition to turbulence}

\begin{figure}
\center
\includegraphics[width=\textwidth]{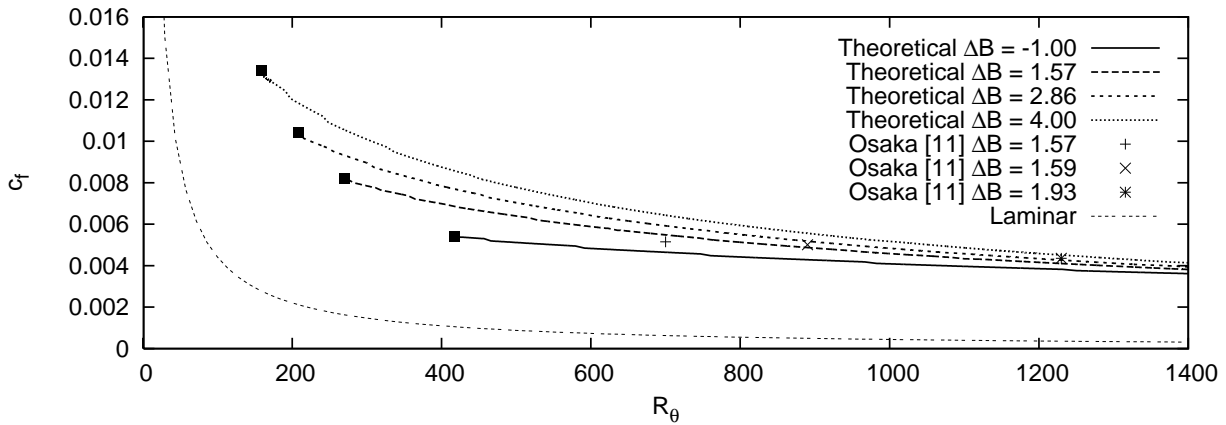}
\caption{Theoretical skin-friction correlation.}
\label{curve}
\end{figure}

The skin-friction law \eqref{eq:cflaw} is shown in Fig.~\ref{curve} for
several different values of roughness function, $\Delta B = \Delta
u/u_{\tau}$. At the critical point where $R_\theta$ is at the minimum
and $c_f$ is at the maximal value, the second and third inflection
points of the velocity profile collide, and the profile will then
only have one inflection point for $R_\theta \leq
R_\theta^\mathrm{crit}$. Therefore, the model indicates the minimal
value of $R_\theta$, i.e. $R_\theta^\mathrm{crit}$, for which a
velocity profile can still be turbulent, i.e. the condition (i) is satisfied.
One can interpret $R_\theta^\mathrm{crit}$ as the minimal Reynolds number
where the transition to turbulence may occur, often called the transitional
Reynolds number in the literature.

\begin{figure}
\begin{minipage}{0.49\textwidth}
\center
\includegraphics[width=\textwidth]{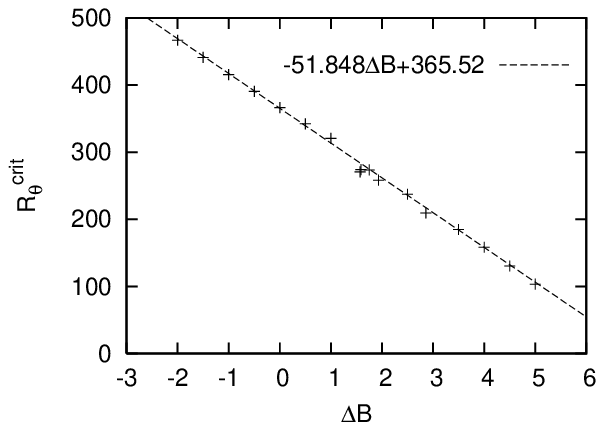}
\caption{Critical value of the Reynolds number.} 
\label{rt}
\end{minipage}
\begin{minipage}{0.49\textwidth}
\center
\includegraphics[width=\textwidth]{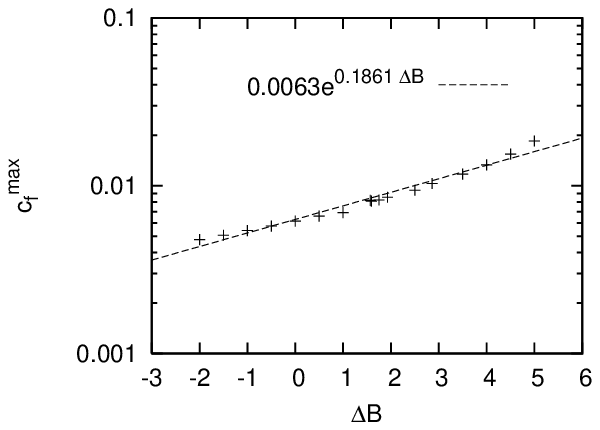}
\caption{Maximum of the skin-friction coefficient.} 
\label{cf}
\end{minipage}
\hfill

\end{figure}

Fig.~\ref{rt} shows how this critical Reynolds number depends on 
the roughness function, which was obtained by computing
$R_\theta^\mathrm{crit}$ for sixteen different values of $\Delta B$. 
The following linear correlation was found:
\begin{equation} \label{eq:Rmin}
R_\theta^\mathrm{crit} = -51.8 \Delta B + 365.5.
\end{equation}

Let us now denote $c_f^\mathrm{max} = f(R_\theta^\mathrm{crit}, \Delta B)$. Note that
$c_f^\mathrm{max}$ is the largest possible value of the skin-friction coefficient
in the turbulent boundary layer.
As demonstrated in Fig.~\ref{curve}, higher values of roughness function will
allow velocity profiles to remain turbulent at lower Reynolds
numbers, but furthermore will result in higher values of
$c_f^\mathrm{max}$. Fig.~\ref{cf}, obtained by evaluating
$c_f^\mathrm{max} = f(R_\theta^\mathrm{crit}, \Delta B)$,
shows how the roughness function is predicted to influence
the maximal value of the skin-friction coefficient. The following correlation was found:
\[
c_{f}^\mathrm{max} = 0.0063e^{0.1861 \Delta B}.
\]

\section{Conclusion}
Based on the Leray-$\alpha$ model of fluid turbulence, a generalized
Blasius equation was formulated to describe streamwise velocity
profiles in turbulent boundary layers with zero pressure gradients.
Solutions of this fifth-order differential equation satisfying a weak
formulation of the von K\'{a}rm\'{a}n log law form a two-parameter
family. The two parameters are the Reynolds number based on
momentum thickness $R_\theta$ and the roughness function $\Delta B$.
This leads to a skin friction correlation $c_f(R_\theta, \Delta B)$ and
predictions of the critical Reynolds number
$R_\theta^\mathrm{crot}(\Delta B)$ and the maximal value of the
skin-friction coefficient $c^{\mathrm{max}}_f(\Delta B)$. The critical Reynolds number is
the minimal value of the transitional Reynolds number (which also
depends on the intensity of the free-stream turbulence).
In particular, it was shown that
the greater $\Delta B$ is, the earlier the transition to turbulence may
occur, and furthermore
the higher the skin-friction coefficient may peak.

\begin{acknowledgments}
We would like to thank Dr. Shinsuke Mochizuki who has kindly provided
us with the experimental data used in this study.
\end{acknowledgments}



\end{document}